\documentclass[twocolumn,prl,showpacs]{revtex4}
\usepackage{graphicx}
\usepackage{dcolumn}
\usepackage{bm}
\usepackage{amsmath}

\begin{document}

\preprint{}
\title{Anomalous Meissner effect in NS junction with spin-active interface}
\author{Takehito Yokoyama$^{1}$, Yukio Tanaka$^{2}$, and Naoto Nagaosa$^{3,4}$}
\affiliation{$^1$Department of Physics, Tokyo Institute of Technology, Tokyo 152-8551, Japan  \\
$^2$Department of Applied Physics, Nagoya University, Nagoya, 464-8603, Japan \\$^3$Department of Applied Physics, University of Tokyo, Tokyo 113-8656, Japan\\
$^4$Cross-Correlated Materials Research Group (CMRG) and Correlated Electron Research Group (CERG), ASI, RIKEN, WAKO 351-0198, Japan 
}
\date{\today}

\begin{abstract}
We investigate Meissner effect in normal metal/superconductor junctions where the interface is spin-active. 
We find that  orbital magnetic susceptibility of the normal metal shows highly nontrivial behaviors. In particular, the magnetic susceptibility depends on the temperature in an oscillatory fashion, accompanied by its sign  change. Correspondingly, magnetic field and current density can spatially oscillate in the normal metal. The possible spontaneous formation of the current pattern is also discussed. 
These results are attributed to the generation of odd-frequency pairing due to the spin-active interface.

\end{abstract}

\maketitle

%--- title ---

%--- author ---

%
%--- address ---

%
%--- date ---

%   Abstract
%-----------------------------------------------------------

%-----------------------------------------------------------

%\section{Introduction}
%-----------------------------------------------------------
Interface phenomena related to the superconductivity constitute a rich field of condensed matter physics. When superconductor is attached to normal metal, Cooper pairs penetrate into the normal metal which acquires superconducting correlation. This is called the proximity effect. As a result, for example, the normal metal has a gap in the density of states or shows Meissner effect\cite{Zaikin,Narikiyo,Higashitani,Belzig,Belzig2}.
In most cases, when lowering temperature, the proximity effect and hence the Meissner response  become stronger. 

However, unexpected behavoir of the proximity induced  Meissner response has been reported:  Mota et al. have uncovered a low-temperature anomaly in the magnetic response of cylindrical structures. At very low temperatures, the susceptibility shows a reentrant behavior and even has paramagnetic region.\cite{Mota,Visani,Mota2,Bernd,Bernd2,Bernd3,Belzig3} However, the origin of this phenomenon still remains unclear. 

Recently, it has been clarified that in normal metal/superconductor junctions, if the interface is spin-active,  induced superconducting pairing in the normal region can change its symmetry from even-frequency pairing to odd-frequency pairing.\cite{Linder,Linder2} Here, even- or odd-frequency means that Cooper pair wavefunction is even or odd with respect to Matsubara frequency (or imaginary time).\cite{Berezinskii} 
If proximity induced pairing symmetry changes, the associated Meissner effect will also change qualitatively. This is the problem we address in this paper.

In this paper, we study Meissner response in the normal metal attached to superconductor where the interface is spin-active. 
We find that orbital magnetic susceptibility of the normal metal shows quite complex dependence on junction parameters. In particular, the magnetic susceptibility depends on the temperature \textit{in an oscillatory fashion, accompanied by its sign  change}. We also show the behavior of magnetic field and current density. These can spatially oscillate in the normal metal. 
These results are attributed to the generation of odd-frequency pairing which stems from the spin-active interface.

%%%%%%%%%%%%%%%%%%%%%%%%%%%%%%%%%%%%%%%%%%%%%%%%%%%%%%%%%%
% Formulation
%%%%%%%%%%%%%%%%%%%%%%%%%%%%%%%%%%%%%%%%%%%%%%%%%%%%%%%%
\begin{figure}[htb]
\begin{center}
\scalebox{0.8}{
\includegraphics[width=10.50cm,clip]{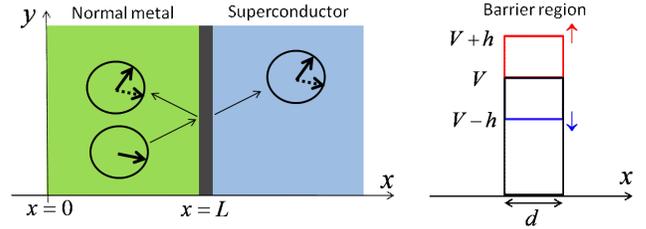}
}
\end{center}
\caption{(Color online) Schematic picture of the model. Left: arrow in the circle represents the direction of spin, which is rotated at the scattering by the spin-active interface. Right: structure of the spin-active interface at $x=L$. 
}
\label{fig1}
\end{figure}

%Let us explain formulation. 
\begin{figure}[htb]
\begin{center}
\scalebox{0.8}{
\includegraphics[width=8.0cm,clip]{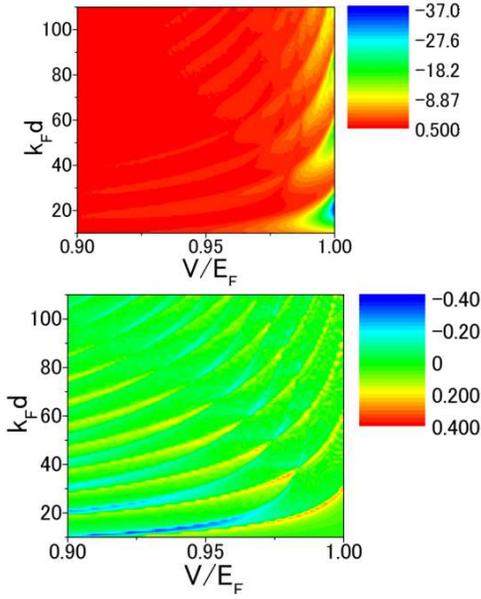}
}
\end{center}
\caption{(Color) Mixing conductances (upper) $G_\phi/G_T$ and (lower) $G_\chi/G_T$ for $h/E_F=0.01$. }
\label{fig5}
\end{figure}

We consider a junction consisting of a diffusive normal metal (DN) with a length $L$ and resistance $R_d$,  and a superconductor. We schematically show the model in Fig. \ref{fig1}.
The interface between the DN and the superconductor at $x=L$ has a resistance $R_b$ (or tunneling conductance $G_T$) and the surface at $x=0$ is specular. A weak external magnetic field $H$ is applied in $z$-direction (perpendicular to the plane of left panel of Fig. 1). 
 We consider spin-active interface at $x=L$ which is described by mixing conductances $G_\phi$ and $G_\chi$ which reflect the spin rotation upon reflection and transmission at the interface, respectively. \cite{Hernando,Cottet}
To evaluate  $G_\phi$ and $G_\chi$, we model magnetic barrier (interface) region as a rectangular potential $V$ with the exchange field $h$ and the width $d$ as shown in the right panel of Fig. \ref{fig1}, following Ref.\cite{Cottet}.
Figure \ref{fig5} shows mixing conductances (upper) $G_\phi/G_T$ and (lower) $G_\chi/G_T$ as functions of $k_F d$ and  $V/E_F$ with  the Fermi wavevector of the DN $k_F$ and the Fermi energy $E_F$. Mixing conductances oscillate with these parameters and $G_\phi/G_T$ rapidly increases when approaching $V/E_F=1$.

To study the Meissner response, we adopt the quasiclassical Green's function theory. The normal and anomalous Green functions are parameterized as $g_\sigma=\cos \theta_\sigma$ and $f_\sigma=\sin \theta_\sigma$, respectively, where $\sigma  (=  \pm  =  \uparrow , \downarrow) $ denotes spin. The Usadel equation\cite{Usadel} in the DN reads
\begin{equation}
D\frac{{\partial ^2 }}
{{\partial x^2 }}\theta _\sigma   - 2\omega _n \sin \theta _\sigma   = 0
\end{equation}
where $D$ and $\omega _n$ are diffusion constant and Matsubara frequency, respectively. 
The boundary conditions are determined by the continuity of the matrix current, and are given by 
\begin{equation}
  \frac{\partial }
{{\partial x}}\theta _\sigma   = 0 
\end{equation}
at $x=0$
and 
\begin{eqnarray}
\frac{{R_b }}{{R_d }}L\frac{\partial }{{\partial x}}\theta _\sigma   =  - g\sin \theta _\sigma   + \sigma f\cos \theta _\sigma \nonumber \\ 
  - i\sigma \frac{{G_\phi  }}{{G_T }}\sin \theta _\sigma   + 2i\frac{{G_\chi  }}{{G_T }}f\left( {g\cos \theta _\sigma   + \sigma f\sin \theta _\sigma   - 1} \right)
\end{eqnarray}
at $x=L$\cite{Cottet}.
Here, $g$ and $f$ are bulk Green's functions in the superconductor. Mixing conductances appear when spin symmetry is broken in the interface region.\cite{Cottet} Consequently, spin triplet pairing is generated, which should be odd in frequency in DN because only $s$-wave pairing can survive  impurity scattering, which is the basic assumption of the Usadel equation. \cite{Linder}
%To evaluate  $G_\phi$ and $G_\chi$, we model magnetic barrier region as a rectangular potential $V$ with the exchange field $h$ and the width $d$, following Ref.\cite{Cottet}.

When a magnetic field is applied parallel to the interface,  rich and nontrivial  screening effect occurs. 
%The local penetration depth is given by 
%\begin{equation}
%\frac{{\lambda _0^2 }}{{\lambda _{}^2 (x)}} = \frac{T}{{T_C }}\sum\limits_{\sigma ,\omega _n  > 0} {\sin ^2 \theta _\sigma  \left( x \right)} 
%\end{equation}
%where $1/\lambda _0^2 =16\pi e^2 N\left( 0 \right)D^2$.
%The averaged value of the local penetration depth is given by 
%\begin{equation}
%\frac{1}{{\lambda _{av}^2 }} = \frac{1}{L}\int_0^L {\frac{1}{{\lambda _{}^2 (x)}}dx} .
%\end{equation}
Within the linear response theory, the current distribution flowing in $y$-direction is given by \cite{Narikiyo,Belzig}
\begin{equation}
j(x) =  - 8\pi e^2 N\left( E_F \right)DT\sum\limits_{\omega _n, \sigma }
{\sin ^2 \theta_\sigma \left( x \right)} A\left( x \right), \label{current}
\end{equation}
where $A(x)$, $N(E_F)$ and $T$ denote the vector potential, the density of states at the Fermi energy and the temperature of the system, respectively.
The Maxwell equation reads
\begin{equation}
\frac{{d^2 }}{{dx^2 }}A\left( x \right) =  - 4\pi j\left( x \right).
\end{equation}
The boundary conditions for $A(x)$ are given by
\begin{eqnarray}
 \frac{d}{{dx}}A\left( 0 \right) = H, \qquad  A\left( L \right) = 0,
\end{eqnarray}
where we have neglected the penetration of magnetic fields into the
superconductor by assuming a small penetration depth in superconductor.

Finally, we obtain the expression of the orbital magnetic susceptibility,
\begin{equation}
- 4\pi \chi  = 1 + \frac{{A\left( 0 \right)}}{{HL}}. \label{chi}
\end{equation}
%\xi _N  = \sqrt {\frac{D}{{2\pi T_C }}}$
%In the following, we focus on diamagnetic susceptibility $\chi$ induced by the proximity effect.

%The role of the mixing conductance can be explained as follows. 
%For low transparent interface, we can linearize the equations. Then, we have the analytical solution of the Usadel equation, which gives 
%\begin{widetext}
%\begin{equation}
%\frac{{\lambda _0^2 }}{{\lambda _{}^2 (x)}} = \frac{T}{{T_C }} \sum\limits_{\omega _n } 2f^2 \frac{{({\mathop{\rm sgn}} (\omega _n )g\cosh kL + \gamma _B kL\sinh kL)^2  - (\gamma _\phi  \cosh kL)^2 }}{{\left[ {({\mathop{\rm sgn}} (\omega _n )g\cosh kL + \gamma _B kL\sinh kL)^2  + (\gamma _\phi  \cosh kL)^2 } \right]^2 }}\cosh ^2 kx 
%\end{equation}
%\end{widetext}
%where  we introduced $k = \sqrt {\frac{{2\left| {\omega _n } \right|}}{D}}$, $\gamma _B  = \frac{{R_b }}{{R_d }}$ and $\gamma _\phi   = \frac{{G_\phi  }}{{G_T }} $.
%This indicates that the odd-frequency triplet component induced by $\gamma _\phi$ contributes to $1/\lambda^2$ with the opposite sign. 

We set $h/E_F=0.01$, $R_d/R_b=10$ and $16\pi e^2 N\left( E_F \right)D^2=1000$.
%Below, $k_F$ and $\xi$ denote Fermi wavenumber and superconducting coherence length which are used to normalize $d$ and $L$, respectively. 
Below, $\xi$ and $T_C$ denote the superconducting coherence length and the transition temperature, respectively.
%which is used to normalize $L$. 
In the following, we plot $- 4\pi \chi$ with its magnitude less than unity since $\left| {4\pi \chi } \right| > 1$ state indicates an instability toward some ordering or sublinear dependence on magnetic field due to the breakdown of linear response theory. 
When $- 4\pi \chi > 1$, the permeability $\mu=1+4 \pi \chi$ becomes negative and hence the energy density $B^2/2 \mu$ is unstable at $B=0$.  This suggests that the new ground state with spontaneous current and magnetic field distribution is stabilized. 
Also, correspondingly, plots of magnetic field and current density are restricted to a certain regime of magnitude.

%%%%%%%%%%%%%%%%%%%%%%%%%%%%%% results

\begin{figure}[htb]
\begin{center}
\scalebox{0.8}{
\includegraphics[width=9.0cm,clip]{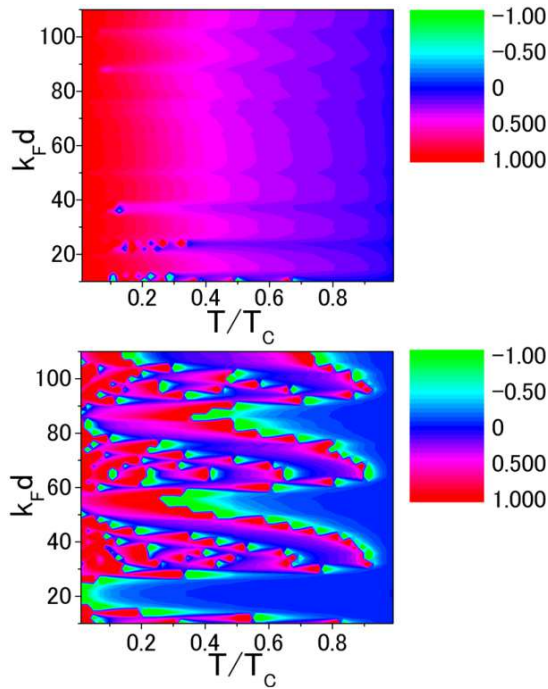}
}
\end{center}
\caption{(Color) Susceptibility $- 4\pi \chi$ at $L/\xi=10$. (upper) $V/E_F=0.95$ and  (lower) $V/E_F=1$. }
\label{fig2}
\end{figure}

\begin{figure}[htb]
\begin{center}
\scalebox{0.8}{
\includegraphics[width=10.0cm,clip]{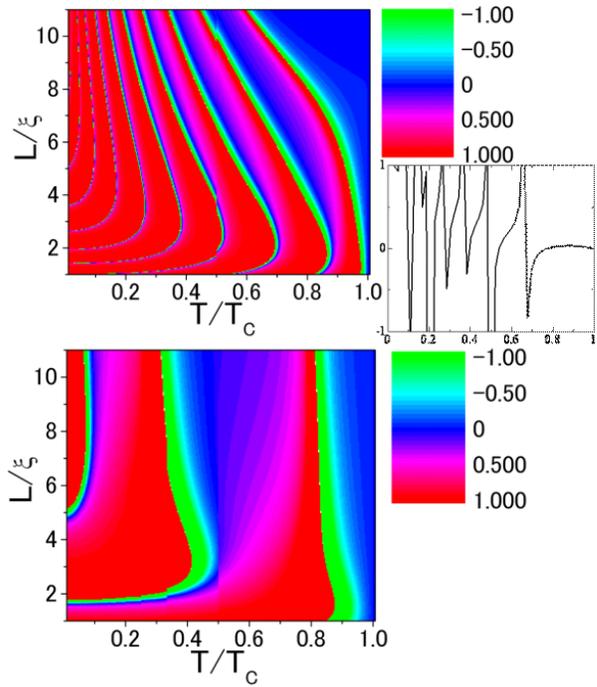}
}
\end{center}
\caption{(Color) Susceptibility $- 4\pi \chi$ at $k_F d=10$. (upper) $V/E_F=0.95$ and (lower) $V/E_F=1$. 
The inset shows susceptibility as a function of $T/T_C$ for $L/\xi=k_F d=10$ and $V/E_F=0.95$. }
\label{fig3}
\end{figure}

\begin{figure}[htb]
\begin{center}
\scalebox{0.8}{
\includegraphics[width=8.0cm,clip]{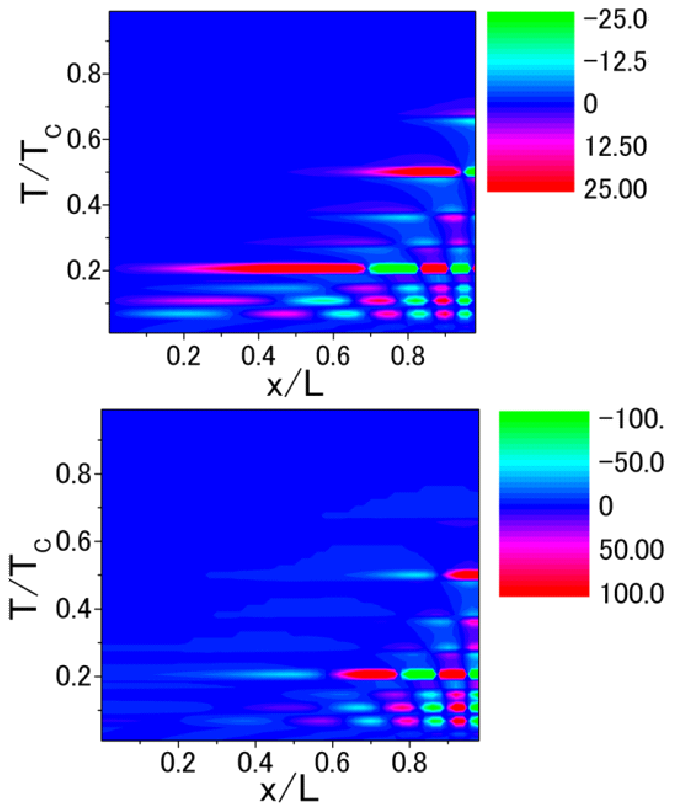}
}
\end{center}
\caption{(Color) Magnetic field (upper) and current density (lower) at $L/\xi=k_F d=10$ and $V/E_F=0.95$. The plots are restricted to a certain regime of magnitude for clarity of figure. }
\label{fig4}
\end{figure}

Figure \ref{fig2} shows susceptibility $-4\pi \chi$ at $L/\xi=10$ with (upper) $V/E_F=0.95$ and  (lower) $V/E_F=1$. At $V/E_F=0.95$, over some region, paramagnetic state, namely that with positive $\chi$, appears. At  $V/E_F=1$, stronger oscillation of the susceptibility accompanied by its sign change is seen. When mixing conductance is present, odd-frequency pairing is generated in the DN region.\cite{Linder,Linder2} This odd-frequency pairing makes it possible to oscillate magnetic field rather than suppress in the DN region\cite{Tanaka}, as expicitly shown below.  Therefore, susceptibility could be positive when odd-frequency pairing correlation is dominant over even-frequency pairing in the DN.

If purely even (odd) frequency pairing state is realized in the DN, $\theta_\sigma$ becomes purely real (imaginary).\cite{Tanaka,Yokoyama}  Then, the sign of screening current Eq.(\ref{current}) depends on whether induced pairing in the DN is even- or odd-frequency pairing. This drastically changes the susceptibility. To understand this qualitatively, let us consider thin limit of DN where spatial dependence of $\theta_\sigma$ is negligible. Then, for purely even-frequency pairing state, 
the Maxwell equation reads
\begin{equation}
\frac{{d^2 }}{{dx^2 }}A\left( x \right) = k^2 A\left( x \right)
\end{equation}
with a real constant $k$. Then, we have
\begin{equation}
A\left( 0 \right) = -\frac{H}{k}\tanh kL.
\end{equation}
Upon insertion of this equation into Eq.(\ref{chi}), we find that $- 4\pi \chi$ is positive definite. On the other hand, for purely odd-frequency pairing state, the Maxwell equation reads
\begin{equation}
\frac{{d^2 }}{{dx^2 }}A\left( x \right) =  - \kappa ^2 A\left( x \right)
\end{equation}
with a real constant $\kappa$. Then, we obtain
\begin{equation}
A\left( 0 \right) = -\frac{H}{\kappa }\tan \kappa L.
\end{equation}
Substituting this equation into Eq.(\ref{chi}), we find that $- 4\pi \chi$ can change its sign. Moreover, it can show divergent behavior near $\kappa L=\pi/2$ mod $\pi$.
%although $\left| {4\pi \chi } \right| > 1$ state is thermodynamically unstable.
In this way, we can understand the behavior of the susceptibility in Figure \ref{fig2}.  Positive sign of $\chi$ means that proximity induced superconductivity shows paramagnetism. 
%This could be interpreted as follows. 
It has been known that in $d$-wave superconductors, paramagnetic contribution to Meissner effect arises from  the Andreev surface states. \cite{Walter,Higashitani2,Barash,Tanaka2} 
%In the system we consider, when odd-frequency pairing is dominant in DN, there is a finite density of states inside the minigap in the DN: rather the density of states has a zero energy peak. \cite{Linder,Yokoyama}
%Thus, the structures of density of states are similar. However, 
In stark contrast, paramagnetic Meissner effect predicted here does not require unconventional superconductivity.

%Therefore, quasiparticle current is possible at low energy, in contrast to conventional even-frequency superconductivity. Hence, this quasiparticle current can contribute to paramagnetic susceptibility. 

Figure \ref{fig3} shows susceptibility at $k_F d=10$ with (upper) $V/E_F=0.95$ and  (lower) $V/E_F=1$. The  susceptibility oscillates with temperature while $L$ dependence is weaker. The inset shows susceptibility as a function of $T/T_C$ for $L/\xi=k_F d=10$ and $V/E_F=0.95$. Reflecting the presence of odd-frequency superconductivity, the susceptibility shows oscillating divergent behavior (The border between red and green regions in the main panel corresponds to the divergence of the susceptibility).

Figure \ref{fig4} depicts normalized magnetic field $H(x)/H$ (upper) and current density (lower) at $L/\xi=k_F d=10$ and $V/E_F=0.95$. 
The current density $j(x)$ is plotted in the unit of $T_C LH/2D$. 
For $T_C \sim 1$meV, $L \sim 1 \mu$m, $H \sim 1$G and $D \sim 10^{-2}$m/s, $T_C LH/2D \sim 10^{10}$A/m$^2$. 
As seen, both  magnetic field and current oscillate in space.  
In a similar way to obtain Eq.(11), we can show that the oscillation is due to the odd-frequency pairing. Therefore, Figure \ref{fig4} also indicates that odd-frequency pairing does not repel magnetic field. 
The relation between susceptibility and magnetic field can be obtained along the same line:
\begin{equation}
 - 4\pi \chi  = 1 - \frac{{\tan \kappa L}}{{\kappa L}} = 1 - \frac{{\sin \kappa L}}{{\kappa L}}\frac{{H(L)}}{H}.
\end{equation}
This indicates that, to realize paramagnetic state, $H(L)$ has to be larger than $H$, namely, magnetic field at the interface should be larger than that at the surface of the DN.

To attain anomalous Meissner effect, dominant odd-frequency pairing in the DN is required. This can be acheived when magnitude of mixing conductance is comparable to that of tunneling conductance ($G_T$)\cite{Linder}. 
%Thus, we study how $G_\phi$ and $G_\chi$ can be large as functions of the width $d$ and the potential $V$ of the barrier region.  
As seen from Figure \ref{fig5}, to enhance effect of spin-activeness, low interface transparency, namely, large potential barrier $V$ is desirable. 
This may be acheived by using ferromagnetic insulator such as  EuO, EuS or La$_2$BaCuO$_5$ as a tunneling barrier.

We find divergent behaviors of the susceptibility, the magnetic field and  the current density. 
However, their magnitudes would be reduced due to spin orbit scattering or magnetic scattering in the sample \cite{Linder3} but the behaviors presented above are  robust against this effect within the reasonable range of $1/(T_C \tau_s) < 1$ with $\tau_s$ being the lifetime of the spin.  

Finally, stability and Meissner effect of odd-frequency pairing in the bulk have been a controversial issue (see Refs. \cite{Solenov,Kusunose}). 
However, the presence of odd-frequency pairing correlation in this paper is due to spin symmetry breaking at the interface and thus irrelevant to these
problems. Namely, the proximity effect is uniquely determined by the boundary conditions.

In summary, we have studied Meissner response in the normal metal attached to superconductor where the interface is spin-active. 
We found that due to the spin-active interface, the magnetic susceptibility shows very complex behaviors. In particular, the susceptibility depends on the temperature in an oscillatory fashion, accompanied by its sign  change. We also showed that magnetic field and current density can spatially oscillate in the normal metal. The possible spontaneous formation of the current pattern was also discussed. 
These results are attributed to the generation of odd-frequency pairing arising from the spin-active interface. 
Our results could be confirmed by experiments with $\mu$-SR or microwave resonance.

The authors thank D. J. Scalapino for useful discussion.  
This work is supported by Grant-in-Aid for Scientific Research
(Grants No. 17071007, No. 17071005, No. 19048008,
No. 19048015, No. 20654030, No. 22103005, No. 22340096 and No. 21244053) from the Ministry of Education, Culture, Sports, Science and Technology of Japan,
Strategic International Cooperative Program (Joint Research
Type) from Japan Science and Technology Agency,
and Funding Program for World-Leading Innovative RD on
Science and Technology (FIRST Program).

%---------------------

\end{document}